\documentclass[sigconf]{acmart}
\usepackage{float}
\usepackage{subfigure}
\AtBeginDocument{%
  \providecommand\BibTeX{{%
    \normalfont B\kern-0.5em{\scshape i\kern-0.25em b}\kern-0.8em\TeX}}}
\usepackage{tabularx}
\usepackage{bbding}
\usepackage{graphicx}
\usepackage{geometry}
\usepackage{subfigure}
\usepackage{amsmath}
\usepackage{makecell}
\usepackage{float}
\usepackage{color}
\usepackage{booktabs}
\usepackage{caption}
\usepackage{tabu}
\usepackage{hyperref}
\usepackage{CJKutf8}

\copyrightyear{2025}
\acmYear{2025}
\setcopyright{acmlicensed}
\acmConference[C\&C '25]{Creativity and Cognition}{June 23--25, 2025}{Virtual, United Kingdom}
\acmBooktitle{Creativity and Cognition (C\&C '25), June 23--25, 2025, Virtual, United Kingdom}
\acmDOI{10.1145/3698061.3726917}
\acmISBN{979-8-4007-1289-0/2025/06}

\author{Sijia Liu}
\authornote{Equal contribution.}
\email{sijialiu7-c@my.cityu.edu.hk}
\orcid{0000-0002-2502-0089}
\affiliation{
\institution{Studio for Narrative Spaces\\City University of Hong Kong}
\city{Hong Kong}
\country{China}}

\author{Xiaoke Zeng}
\authornotemark[1] 
\email{xzeng36@cityu.edu.hk}
\orcid{0009-0007-5987-502X}
\affiliation{
\institution{Studio for Narrative Spaces\\City University of Hong Kong}
\city{Hong Kong}
\country{China}}

\author{Fengyihan Wu}
\email{wfyh2001@163.com}
\orcid{0009-0005-1546-6924}
\affiliation{
\institution{University college London}
\city{London}
\country{United Kingdom}}

\author{Shu Ye}
\email{yeshu935776019@gmail.com}
\orcid{0009-0007-0743-365X}
\affiliation{
\institution{University college London}
\city{London}
\country{United Kingdom}}

\author{Bowen Liu}
\email{bowenliu@cityu.edu.hk}
\orcid{0009-0001-9314-0373}
\affiliation{
\institution{Studio for Narrative Spaces\\City University of Hong Kong}
\city{Hong Kong}
\country{China}}

\author{Sidney Cheung}
\email{sidneycheung@cuhk.edu.hk}
\orcid{0000-0002-0989-3824}
\affiliation{
\institution{Chinese University of Hong Kong}
\city{Hong Kong}
\country{China}}

\author{Richard William Allen}
\email{rwallen@cityu.edu.hk}
\orcid{0000-0003-2826-0990}
\affiliation{
\institution{City University of Hong Kong}
\city{Hong Kong}
\country{China}}

\author{RAY LC}
\authornote{Correspondences should be addressed to LC@raylc.org.}
\email{LC@raylc.org}
\orcid{0000-0001-7310-8790}
\affiliation{
\institution{Studio for Narrative Spaces\\City University of Hong Kong}
\city{Hong Kong}
\country{China}}

\begin{document}


\title[Salt is the Soul of Hakka Baked Chicken]{"Salt is the Soul of Hakka Baked Chicken": Reimagining Traditional Chinese Culinary ICH for Modern Contexts Without Losing Tradition}

\begin{abstract}

Intangible Cultural Heritage (ICH) like traditional culinary practices face increasing pressure to adapt to globalization while maintaining their cultural authenticity. Centuries-old traditions in Chinese cuisine are subject to rapid changes for adaptation to contemporary tastes and dietary preferences. The preservation of these cultural practices requires approaches that can enable ICH practitioners to reimagine and recreate ICH for modern contexts. To address this, we created workshops where experienced practitioners of traditional Chinese cuisine co-created recipes using GenAI tools and realized the dishes. We found that GenAI inspired ICH practitioners to innovate recipes based on traditional workflows for broader audiences and adapt to modern dining contexts. However, GenAI-inspired co-creation posed challenges in maintaining the accuracy of original ICH workflows and preserving traditional flavors in the culinary outcomes. This study offers implications for designing human-AI collaborative processes for safeguarding and enhancing culinary ICH.



\end{abstract}




\maketitle

\section{Introduction}\label{sec:Introduction}

Culinary traditions display regional cultural diversity through specialized techniques and meanings embedded in recipes and cooking practices \cite{lee_cooking_2023, frey_preserving_2015}. However, traditional culinary knowledge is increasingly shaped by modern societal demands, requiring adaptation to ensure its survival \cite{Sustainable_Development, deaton1998economies}. Food-related cultural practices first received official recognition as Intangible Cultural Heritage (ICH) by UNESCO in 2010 \cite{unesco_decision_2010}, and this has since encouraged a rethinking of how culinary heritage can be sustained and recognized globally. Chinese traditional cuisine serves as an example of a rich cultural asset facing such re-evaluation. Although renowned for its historical depth and cultural significance \cite{li2004traditional, cheung_food_2002}, Chinese cuisine remained absent from UNESCO’s inscribed lists for years \cite{noauthor_unesco_nodate, unesco2024lists}, unlike the culinary traditions of neighboring countries such as Korea and Japan. Related research has shown that local ICH culinary practitioners in China often emphasize “authenticity” and “distinctive culture” in their understanding of heritage, whereas the UNESCO Convention also acknowledges the dynamic nature of ICH, including processes of re-making and reshaping cultural practices \cite{demgenski_culinary_2020}. This situation highlights the need for strategic adaptations that safeguard traditional cuisine by not only preserving its authenticity but also enabling its expressive development and evolution to support broader global recognition.

To meet these challenges, the transmission of ICH culinary practices requires innovative approaches that go beyond mere preservation to technology-mediated forms of community engagement. Research has shown that technology plays a key role in the protection and dissemination of ICH \cite{liu_digital_2023, lu_i_2019}. Virtual Reality (VR) \cite{de_paolis_virtual_2022}, Augmented Reality (AR) \cite{bekele_survey_2018}, serious games \cite{mortara_learning_2014, grammatikopoulou_adaptive_2019}, narrative interventions\cite{song_drizzle_2022, lc_designing_2021, lc_designing_2022}, 3D visualizations \cite{selmanovic_improving_2020, comber_designing_2014}, and Generative AI (GenAI) \cite{fu_being_2024, heng_echoes_2024} have shown potential to foster engagement and preserve cultural practices. These technologies are also increasingly used in Human–Food Interaction (HFI) research to document and envision food practices \cite{comber_designing_2014}. However, most current efforts still emphasize documentation and passive engagement for audiences, while the innovation and reinterpretation of ICH in global contexts remain underexplored through technologies.

Among these tools, GenAI in particular has potential to address this gap. Due to its recent advancements in reimagining and visualizing abilities \cite{ko2023large, rombach_high-resolution_2022, weisz_design_2024}, GenAI shows promise in expressively reinterpreting and recreating cultural heritage \cite{fu_being_2024, heng_echoes_2024, he_i_2025}. In addition, HFI research demonstrates how generative models can inspire online recipes and digitally reconstruct traditional cooking methods \cite{LLaVA-Chef, Can_AI_Help}. These capabilities position GenAI as a potential tool for both supporting the continuity and empowering innovative expression of ICH culinary practices. It may help culinary professionals adapt traditional cuisine to the developing new situations during the practice process.

To investigate how to enhance and promote traditional Chinese cuisines as ICH, with the support of GenAI, we organized in-person workshops with participants who are experienced practitioners specializing in traditional Chinese cooking and hold their own insights into culinary culture. By utilizing GenAI, we aimed to assist these practitioners in reimagining traditional cuisine and exploring potential avenues for its recreation. Through this workshop-based study, we attempted to understand how GenAI could help the practitioners to find new possibilities in traditional Chinese cuisine, by generating novel interpretations of traditional recipes, offering suggestions for ingredient substitutions, and visualizing possible adaptations that align with contemporary societal needs. Additionally, the workshops facilitated discussions on how the biases of GenAI tools, the stereotypes assumptions of participants, and gaps in knowledge might limit this co-creation process. 

This study makes the following contributions to the HCI fields through the workshops: 1) It demonstrates the application of GenAI in facilitating innovation within traditional cultural domains, specifically showing how GenAI could assist practitioners in reimagining and adapting traditional Chinese cuisines while maintaining their cultural essence. 2) It provides insights into how Chinese cuisine practitioners could leverage new technologies to balance playful culinary innovation with the imperative of cultural preservation, offering practical examples of how AI can inspire creativity while respecting tradition. 3) It offers implications for designing human-AI co-creative workshops focused on cultural heritage safeguarding. Overall, the study demonstrated how ICH cuisine practitioners used GenAI tools to recreate recipes, adapting to global developments while striking a balance between cultural innovation and preservation.

\section{Background}\label{sec:Background}
\subsection{Food as Intangible Cultural Heritage}
ICH (Intangible Cultural Heritage) was officially defined by UNESCO in 2003 as traditions and knowledge passed down through generations, forming a vital part of cultural identity \cite{unesco_convention_2003}. Food-related culture was first included in 2010 on UNESCO's annually updated list of ICH practices in need of safeguarding and promotion \cite{unesco2024lists, unesco_decision_2010}. Since then, the recognition of food heritage has gained global attention in both academic and societal contexts \cite{romagnoli_gastronomic_2019, herman_food_2024, Csergo_2018}. As of 2024, nearly 40 culinary traditions worldwide have been inscribed on the UNESCO list \cite{unesco2024lists}. 

However, like other forms of ICH, culinary traditions face ongoing challenges in preservation and protection. Variations in household-specific cooking methods, diverse family recipes \cite{frey_preserving_2015, dietz_preserving_2017}, and reliance on oral or manuscript-based transmission of techniques \cite{unesco_practices_2023, frey_preserving_2015, dietz_preserving_2017} all complicate the safeguarding and promotion of ICH culinary heritage.

In contemporary contexts, the transmission of culinary practices involves even more complex issues. Beyond their cultural value, food traditions are closely tied to health, society, and the economy \cite{li2004traditional}. They are also influenced by local agriculture, tourism, and economic development \cite{unesco_traditional_2010}. These connections introduce new challenges to ICH culinary development, such as the scarcity of traditional ingredients \cite{reuter_endangered_2022}, the shift toward modern cooking equipment \cite{sivakaranjournal}, the influence of regional tourism \cite{ctx46153656340003408}, evolving family dynamics \cite{olayemi2012effects}, and changing health standards \cite{trichopoulou2007traditional, Broude_2015}. For ICH culinary practices to survive, they must not only be preserved but also adapted and improved to meet changing societal needs \cite{Sustainable_Development, deaton1998economies}.

\subsection{Adapting Chinese ICH Food to a Globalized World}
Chinese cuisine, as a long-standing food tradition, has demonstrated remarkable adaptability throughout history. During the migration of Chinese communities around the world, Chinese cuisine was often improvised and reinvented based on available ingredients and local tastes \cite{cheung2014globalisation}. Today, we can observe the incorporation of nontraditional elements into urban Chinese food, reflecting its dynamic response to globalization \cite{bai2014dietary}. However, by the end of 2024, UNESCO had inscribed 44 Chinese cultural practices on its Intangible Cultural Heritage (ICH) list, but only the traditional tea processing techniques represent China among the approximately 40 food and drink-related entries worldwide \cite{people2023heritage, chinadaily2023spring, unesco2024lists}.

To explore the reasons behind this underrepresentation, we can look to the example of Japan’s traditional dietary culture, Washoku, which was inscribed on UNESCO’s ICH list in 2013 \cite{unesco2024lists}. Local studies reveal that the process of transforming food traditions into recognized ICH often involves various forms of adaptation \cite{Japan_Washoku}. Japanese chefs and food advocates have sought to strengthen local food practices while simultaneously embracing globalization \cite{Global_Engagement}. They believe that reaching a broader international audience through a distinct national cuisine does not compromise cultural identity. As some ICH researchers have noted, this perspective contrasts with that of many Chinese ICH culinary practitioners, who often emphasize traditions and authenticity over reinterpretation, despite the fact that these culinary practices are constantly evolving and adapting \cite{demgenski_culinary_2020}.

For further global recognition and effective safeguarding, Chinese culinary heritage should find a balance between innovation and tradition. As a reference point for this study, adapting Chinese culinary practices to global contexts should involve enhancement and creative evolution while remaining rooted in cultural origins.

\subsection{The Role of Digital Technologies in Safeguarding ICH}
Digital technologies play an increasingly important role in safeguarding ICH, driven by advancements in computational heritage and Information and Communications Technology (ICT) applications. These technologies enable the creation of diverse resources that make ICH more accessible to the public \cite{liu_digital_2023, hou_digitizing_2022}. Digital social media platforms further contribute to the preservation and promotion of ICH. HCI researchers have explored how livestreaming can safeguard ICH practices by enabling real-time demonstration, documentation, and interaction between practitioners and audiences \cite{lu_live_2019, lu_you_2018, lu_i_2019}. Short video platforms have opened up new opportunities for innovation and communication, fostering collaboration and dialogue between ICH practitioners and global viewers \cite{wang_critical_2024}. Web-based and VR-enabled 360-degree reconstructions allow visitors to explore cultural heritage sites in immersive 3D environments \cite{selmanovic_improving_2020, de_paolis_virtual_2022}. Similarly, augmented reality (AR) and mixed reality (MR) are being applied to enhance the visual accessibility and engagement with ICH practices \cite{bekele_survey_2018}. In addition, tangible user interfaces have been designed to preserve indigenous knowledge by encouraging community engagement and interaction \cite{dimitropoulos_capturing_2024}. Serious games have also been developed to raise public awareness of both tangible and intangible cultural heritage, combining education with entertainment \cite{mortara_learning_2014, grammatikopoulou_adaptive_2019}.

In recent years, Generative AI (GenAI), powered by large language models (LLMs), has emerged as a transformative tool in ICH preservation \cite{ko2023large, rombach_high-resolution_2022, weisz_design_2024}. Humans are increasingly collaborating with GenAI tools to support the translation and preservation of unclear or vague cultural materials \cite{mahdavi_goloujeh_is_2024, verheijden_collaborative_2023, fan_contextcam_2024}. Moreover, GenAI has been utilized for co-creation purposes, such as reconstructing and visualizing cultural heritage artifacts for exhibitions \cite{fu_being_2024, lc_present_2024, lc_human_2023, lc_together_2023}. It is also applied in interactive systems to enhance public understanding of ICH by generating motion-based and visualized representations of cultural practices \cite{heng_echoes_2024}.

The integration of digital technologies highlights the potential for innovation in safeguarding ICH while ensuring its accessibility and adaptability in a rapidly changing, globalized world.

\subsection{Human Food Interaction}
The growing interest in food and taste in the HCI field \cite{berkholz_becoming_2023}, leading to the gradually attraction of Human-Food Interaction (HFI) \cite{comber_designing_2014}. HFI researchers conduct practical studies and workshops to explore daily food experiences \cite{choi_food_2012, comber_food_2012, gayler_sensory_2021} as well as envision future sustainable food cultures \cite{deng_future_2021, lindrup_one_2021}.

Technology could play a role in enriching and transforming food-related practices \cite{khot_human-food_2019}. Immersive technologies such as VR, AR, and the Metaverse are being leveraged to create interactive and imaginative food experiences, helping users envision the future of food and its cultural implications \cite{covaci_no_2023}. Similarly, AI is increasingly integrated into the food design process, enabling the creation of innovative and personalized culinary concepts \cite{al-sarayreh_inverse_2023}.

Innovative designs in HFI research also foster playful, social engagement with food. For example, digital creators have designed games to document and disseminate ICH cuisines, making historical and cultural food practices more accessible to wider audiences \cite{cui_protection_2021}. Interactive food exhibitions are another rapidly growing area, blending technology with culinary arts to provide engaging experiences that contribute to the development of the food industry \cite{ji_research_2019}.

Beyond these examples, other prominent topics in HFI research include computational approaches to food and dining experiences \cite{deng_dancing_2023}, food journaling for health and self-discovery \cite{m_silva_investigating_2021}, and the influence of food-related social media content creators on public perceptions and culinary trends \cite{weber_its_2021}. These diverse research directions highlight how technology is reshaping the way we interact with, understand, and celebrate food in both personal and cultural contexts.

Although digital technologies have advanced ICH safeguarding and food-related research, the intersection of ICH food traditions and HFI remains underexplored, particularly in terms of its potential for innovation. Our study addressed this gap by designing participatory workshops that leverage GenAI to help practitioners reimagine, preserve, and enhance ICH food practices, adapting to the modern global culture and fostering both cultural sustainability and creative engagement.

\section{Methods}\label{sec:Methods}
To explore how Chinese cuisine practitioners adapt their culinary processes with GenAI in response to changing times, we conducted individual workshops in participants' professional cooking environments. Fifteen experienced practitioners were recruited for this study. Each participant was asked to select a cuisine and prepare to cook it during the workshop. GenAI tools were then applied to support their recipe recreation.

\begin{figure*}[htbp]
    \centering
    \includegraphics[width=\textwidth]{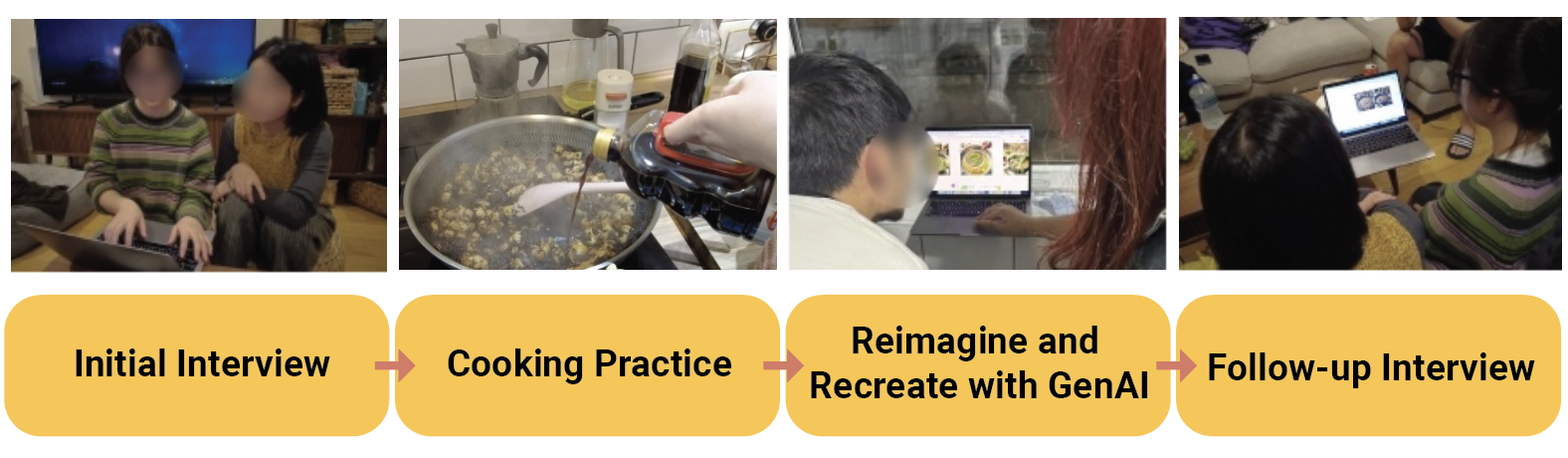}
    \caption{Workshop Structure}
    \label{fig:enter-label}
\end{figure*}

\subsection{Participants Recruitment}
The basic criteria for selecting participants was individuals who identified themselves as Chinese, and had a minimum of six years of experience in preparing Chinese ICH cuisines. The judgment of ICH cuisines based on the official list of ICH recognized by relevant Chinese ICH departments at all levels, including provincial, municipal, and prefectural ones \cite{ihchina_project, shidian2023}. These cuisines, recognized for their authenticity and cultural significance, were ideal for exploring the recreation and enhancement of ICH culinary processes. 

We used convenience sampling to select participants who were accessible and willing to participate \cite{etikan2016comparison}. To identify and recruit qualified participants, we employed three recruitment strategies: (1) engaging researchers' personal networks to connect with culinary practitioners, (2) utilizing academic networks, such as culinary culture researchers who had access to experienced chefs, and (3) directly reaching out to cooks at restaurants. These practical strategies enabled us to recruit amenable participants with substantial experience in preparing Chinese cuisine.

Finally, we selected fifteen participants who met our criteria. Eight among the participants were professional chefs and seven were amateur cooks. Their culinary expertise spanned different regions of China. For amateur cooks, we not only required them to meet our basic criterion of a minimum number of years of experience. We also considered the types of cuisine they specialized in, their learning experiences with the cuisine, and records of their cooking process or finished dishes. We primarily selected amateurs who cooked the cuisine of their hometown or the region where they grew up, learning their cooking practices from family recipes. This form of culinary transmission is similar to the way intangible cultural heritage is passed down through generations, preserving traditional cooking knowledge and techniques through oral methods \cite{frey_preserving_2015, dietz_preserving_2017}.  The detailed demographic information of the participants is presented in Table\ref{tab:table1}.

\raggedbottom
\begin{table*}[htbp]
\centering
\caption {\label{tab:table1} Demographic Information of Participants}
\small 
\begin{tabular} {p{0.3cm}p{4cm}p{1cm}p{0.5cm}p{1cm}p{5.5cm}} 
\toprule
\textbf{ID} & \textbf{Dish} & \textbf{Gender} & \textbf{Age} &   \textbf{Occupation} & \textbf{Experience of Cooking} \\

\midrule
$P1$ & Hakka-style Stuffed Tofu & Male & 29 & Amateur & 10 years of Hakka cuisine \\
$P2$ & Claypot Rice & Male & 56 & Chef & 30 years of Cantonese cuisine \\
$P3$ & Basil Stir-fried Clams & Male & 26 & Amateur & 6  years of Cantonese cuisine \\
$P4$ & Hunan Style Stir-fried Beef & Male & 50 & Chef & 30 years of Hunan cuisine \\
$P5$ & Garlic Vermicelli Shrimp Pot & Female & 30 & Amateur & 10 years of Sichuan and Chongqing cuisine \\
$P6$ & Juicy Soup Buns & Male & 27 & Chef & 7 years of Dim Sum \\
$P7$ & Liuyang Steamed Pork Ribs & Female & 25 & Amateur & 6 years of Jiangxi Cuisine \\
$P8$ & Yu-Shiang Shredded Pork & Male & 25 & Amateur & 6 years of Henan cuisine \\
$P9$ & Stewed Pork with Preserved Vegetables & Female & 27 & Amateur & 6 years of Jiangsu and Zhejiang cuisine \\
$P10$ & Guiyang Fermented Soybean Hot Pot & Male & 43 & Chef & 10 years of Sichuan cuisine, Guizhou cuisine \\
$P11$ & Shaanxi Oil-spilled Noodles & Female & 32 & Chef & 8 years of Shaanxi cuisine, Fusion French Cuisine \\
$P12$ & Wonton Noodles & Female & 31 & Amateur & 10 years of Cantonese cuisine \\
$P13$ & Zhaoqing Zongzi & Female & 58 & Amateur & 30 years of Cantonese cuisine \\
$P14$ & Chaoshan Beef Balls & Male & 40 & Chef & 20 years of Chaoshan cuisine \\
$P15$ & Hakka Salt-baked Chicken & Male & 55 & Chef & 30 years of Cantonese cuisine \\

\bottomrule
 \end{tabular}
\end{table*}

In addition, most participants had no experience working with GenAI, except for P1, P7, and P8, who had a few months of exposure to ChatGPT and Chinese local LLMs. 

\subsection{Workshop Procedure}
We mainly followed four steps in the workshop structure (Fig.\ref{fig:enter-label}).
\subsubsection{Initial Interview and Guidance} At the start of each workshop, we conducted interviews with participants regarding their traditional cooking methods and the specific details of their chosen cuisine. We asked about their cooking backgrounds, the specific ICH dish they had selected for the workshop, as well as history, culture and unique characteristics of the selected dishes. We then guided participants to discuss adaptations in cooking practices according to their experience, including considerations for different target audiences (tasters), balance between taste and health, ingredient substitutions, visual style, recipe modifications, cooking time and kitchen layout, and cooking utensils, etc. We also introduced selected stories of UNESCO-inscribed culinary traditions to provide cultural references that helped participants better understand culinary adaptation in contemporary global contexts.

\subsubsection{Cooking Practice} After the initial interview, the participants prepared their selected dishes. They were asked to prepare step by step, follow their recipe, and avoid any omissions or improvisations. During the process, researchers recorded videos and took detailed notes to capture critical information for later comparison with the recreated recipes if needed.

\subsubsection{Reimagine and Recreate with GenAI} After completion of the cooking, we guided participants to generate reimagined culinary images and recipes using GenAI tools, drawing from inspiration about culinary evolution and adaptations discussed in the earlier interview. We encouraged them to explore new possibilities for their selected ICH dishes while interacting with GenAI, particularly by imagining how these cuisines might adapt to different situations and challenges, such as international cooking competitions, health-conscious adaptations, or incorporation of different culinary techniques. They were also encouraged to incorporate ideas they had long considered but had not previously attempted, while being given the creative freedom to innovate the ICH cuisine according to their preferences. 

We selected ChatGPT as our primary tool due to its advanced natural language understanding and text-image generation capabilities \cite{openai_introducing_nodate}. Additionally, during the recruitment process, several participants (P1, P7, and P8) with prior experience using GenAI expressed concerns about cultural bias in ChatGPT, particularly in relation to Chinese contexts. Based on their suggestions, we incorporated two LLMs developed by Chinese corporations, Doubao and Zhipu into the workshop process \cite{doubao2024vision, zhipu2024}. We guided all participants to use ChatGPT first but informed them that they could switch to Doubao or Zhipu if they wished to.

Since most other participants had no experience working with GenAI, we first introduced them to the basics of using GenAI, including how to ask questions and generate text and image content. We then guided participants to collaborate with GenAI tools to generate updated recipes and food images. 

Participants were enabled to ask their own questions and upload their own recipes and images as references throughout the process. They iteratively refined prompts until the generated content met their creative expectations. During this process, participants were also allowed to use the think-aloud protocol to verbalize their thoughts and reflections \cite{Direct_from_the_source, The_think_aloud_method}, which were audio-recorded. Throughout their co-creation process, we remained available to provide technical support whenever needed. 

If the participants wanted to reprepare the dishes based on the AI-generated recipes, it would be also encouraged and recorded.

\subsubsection{Follow-up Interview} Following the recreation phase, we conducted semi-structured interviews using the photo-elicitation method \cite{harper_meaning_1986,harper_talking_2002}. We gathered feedback on multiple aspects, including participants' responses to AI-generated content, evaluation of the overall cooking and interaction processes, knowledge gained, memories formed, and their experiences using AI.

\subsection{Data Collection and Analysis}
Throughout the workshop, we recorded audio, video, and took detailed field notes with the participants' consent. Additionally, we preserved the prompts used in their interactions with GenAI. All audio recordings were transcribed verbatim into text transcripts, and videos of cooking processes were reviewed multiple times to complement and verify our field notes. All field notes, think-aloud data, and interview transcripts were collected for obtaining qualitative codes.

These data were initially transcribed and then independently open-coded by four researchers following the principles of thematic analysis \cite{braun_using_2006, thomas_general_2006, wicks_coding_2017}. Multiple rounds of iterative discussions and coding were conducted until the researchers reached consensus on the final themes.

\section{Results}\label{sec:Results}
\subsection{Innovative Approaches to Cuisine Co-Created with GenAI}
According to our analysis of the data, the participating chefs’ approaches to recreating cuisine with GenAI can be categorized into three main themes: modifications to traditional dishes, enhancing the health value of the cuisine, and incorporating elements from other culinary cultures.

\subsubsection{New Possibilities of Traditional Dishes}
\label{4.1.1}
\begin{figure*}
    \centering
    \includegraphics[width=1\linewidth]{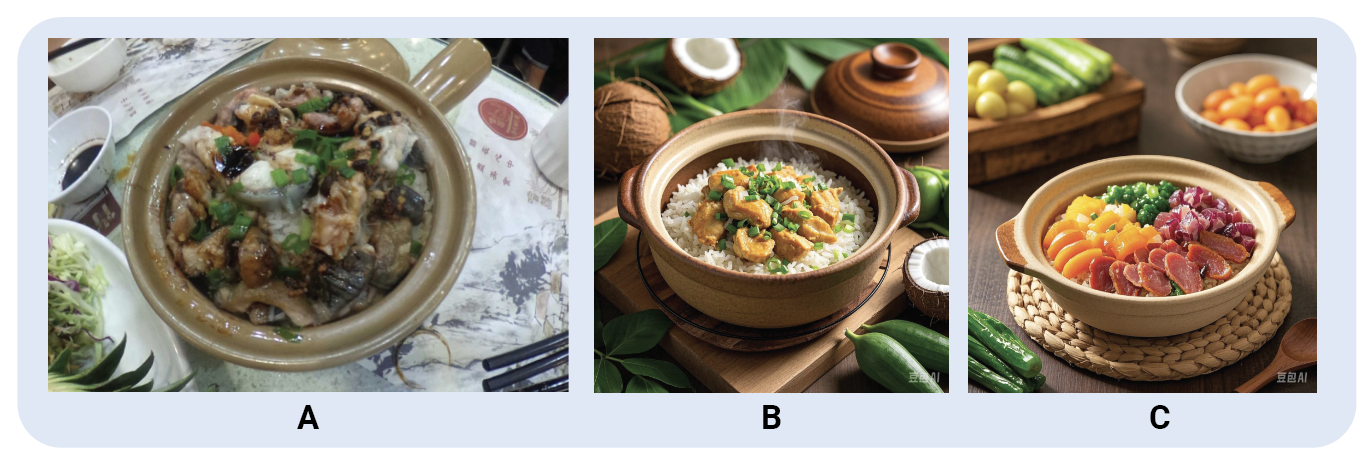}
        \caption{A) Claypot Rice prepared by P2; B) Thai Coconut Chicken Fusion version of Claypot Rice generated by GenAI; C) Fruit-inspired version of Claypot Rice generated by GenAI.}
    \label{fig:p2}
\end{figure*}

During the co-creation with GenAI, participating chefs considered making modifications to traditional dishes, such as improving cooking methods or incorporating new ingredients, to make the dishes more appealing to a wider audience. For instance, P2 attempted to generate variations of Claypot Rice suited to different regional tastes, exploring possibilities to attract more people to enjoy this dish (Fig.\ref{fig:p2} B) and C)). Similarly, P1 inquired about a simpler way to prepare Hakka-style Stuffed Tofu, aiming to create an easy version for home cooks who love the dish. These innovative ideas stem from a shared focus on promoting ICH cuisines.

\begin{figure*}
    \centering
    \includegraphics[width=1\linewidth]{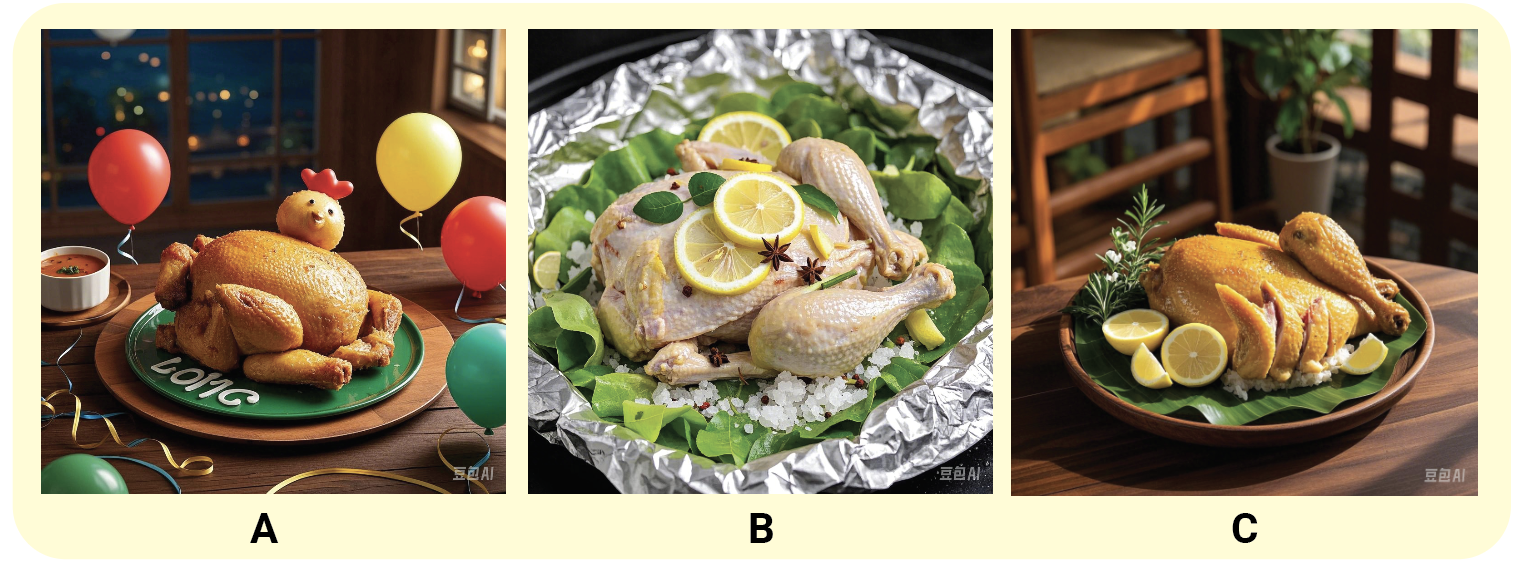}
        \caption{A) Innovated Hakka Salt-Baked Chicken for youth gatherings generated by GenAI; B) Low-calorie version of Hakka Salt-Baked Chicken innovated by GenAI; C) Thai cuisine fusion version of Hakka Salt-Baked Chicken innovated by GenAI.}
    \label{fig:p15}
\end{figure*}

P15 explored new possibilities for Hakka Salt-Baked Chicken that potentially popular among young generation, and GenAI suggested a crispy skin variation (Fig.\ref{fig:p15} A)). P15 remarked, "This version is indeed more suitable for young people gathering situations, offering a novel look and taste. It has greater potential appeal to younger audiences, helping to establish this ICH dish in the hearts of the younger generation and ensuring its continued legacy."

Similarly, P13 tried to find new ideas for Zhaoqing Zongzi, with GenAI proposing a filling that includes stretchy cheese. P13 commented, "From an innovation perspective, this breaks through the limitations of traditional thinking, boldly incorporating novel ingredients and popular flavors into ICH dishes. It introduces combinations that were previously hard to imagine within traditional practices, giving the dish a fresh and intriguing look. This approach significantly helps attract younger audiences and expand the appeal of ICH dishes to those with different flavor preferences."

\subsubsection{Health-Enhanced Variations of Cuisine}
\label{4.1.2}
Increasing the health value of dishes is another factor participants aimed to improve, as they believed aligning with modern nutritional balance standards would make the dishes more widely accepted by diverse groups. 

\begin{figure*}
    \centering
    \includegraphics[width=1\linewidth]{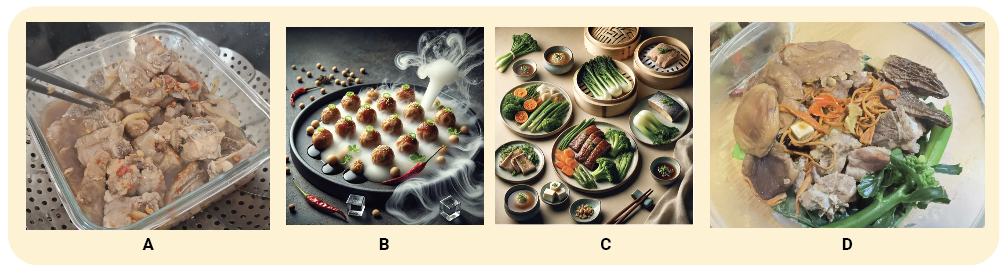}
        \caption{A) Liuyang Steamed Pork Ribs prepared by P7; B) Liuyang Steamed Pork Ribs multi-dish health set meal innovated by GenAI; C) Molecular gastronomy version of Liuyang Steamed Pork Ribs innovated by GenAI; D) Healthier version of Liuyang Steamed Pork Ribs with added vegetables prepared by P7, based on B) and the related AI-generated recipe.}
    \label{fig:p7}
\end{figure*}

P7 considered creating a multi-dish set meal inspired by their expertise in Liuyang Steamed Cuisine (Fig.\ref{fig:p7} B)). GenAI suggested that, in addition to Liuyang Steamed Pork Ribs as the main dish, complementary dishes like steamed fish with scallion sauce and steamed tofu could be included as high-protein additions.

Similarly, an innovative version of Hakka-style Stuffed Tofu generated by GenAI included tomatoes and beans as side garnishes. P1 recognized this imagination, noting that the addition of vegetables not only enhanced the visual appeal of the dish but also increased its nutritional value.

P15 wanted to make Hakka Salt-Baked Chicken more suitable for young fitness enthusiasts  (Fig.\ref{fig:p15} B). GenAI proposed a low-calorie, fat-reduced version by replacing certain ingredients or reducing the use of original condiments. P15 commented, "This version aligns with people's current pursuit of healthy eating. By modifying the traditional recipe, it allows fitness enthusiasts and those focused on body management to enjoy the deliciousness of salt-baked chicken without any guilt. This undoubtedly broadens the audience for our ICH dish."

\subsubsection{Fusion with Other Culinary Traditions}
\label{4.1.3}
Some participating chefs believe that combining Chinese cuisine with other culinary traditions can be an innovative way to introduce ICH dishes to international audiences. For example, P8 collaborated with GenAI to create the Yu-Shiang Shredded Pork Burger, blending traditional Eastern flavors with a Western dish format (Fig.\ref{fig:p8} B)).

\begin{figure*}
    \centering
    \includegraphics[width=1\linewidth]{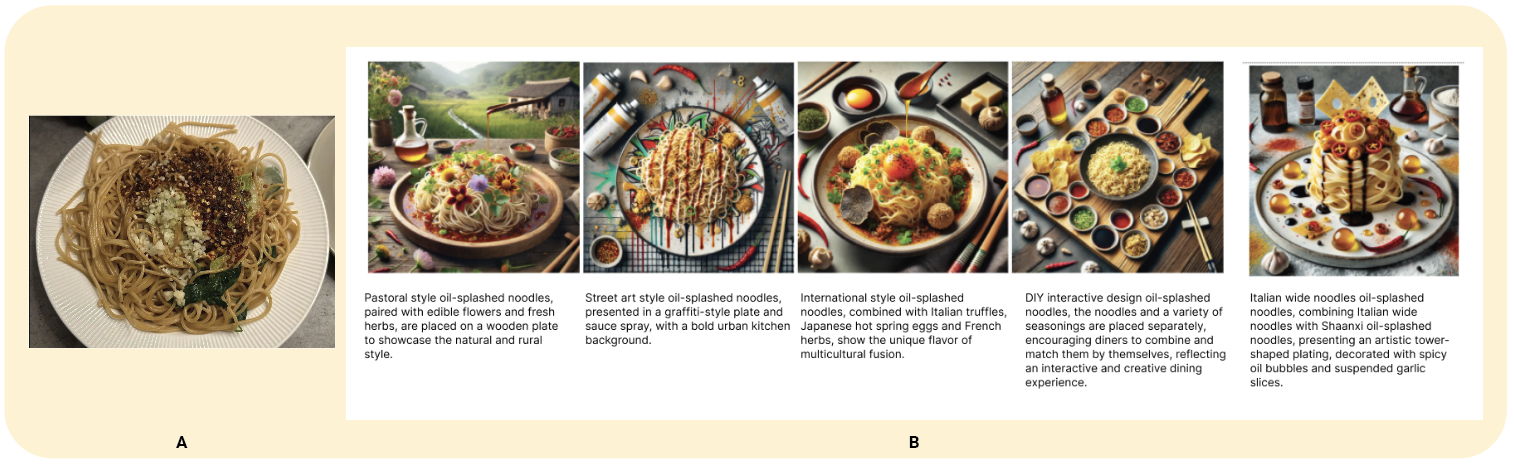}
        \caption{P11: A) Shaanxi Oil-Spilled Noodles cooked by P11 B) A series of AI-generated fusion dishes blending Shaanxi Oil-Spilled Noodles with elements of Western cuisine.}
    \label{fig:p11}
\end{figure*}

P11, who already has experience in preparing both Shaanxi cuisine and French fusion dishes, explored more ideas during the workshop. GenAI provided her several suggestions for fusing Shaanxi Oil-Spilled Noodles with French cuisine  (Fig.\ref{fig:p11} B)). P11 shared, "Western cuisine and Shaanxi food are two seemingly unrelated styles, but AI tools gave me a lot of inspiration, especially on how to use local ingredients and French plating habits to merge Shaanxi flavors. It opened up my perspective—for instance, pairing local cheese with chili oil, adding honey to garnish Shaanxi's cold noodles, or even reinterpreting oil-spilled noodles using French molecular gastronomy techniques."

GenAI also suggested a new version of Hakka Salt-Baked Chicken by incorporating elements of Thai cuisine for P15  (Fig.\ref{fig:p11} C)). P15 appreciated this innovative recipe, saying, "The Thai-inspired version cleverly integrates the unique flavors of Southeast Asia. If promoted well, it might help our Salt-Baked Chicken cross borders and become loved by people from different countries and regions."

\subsection{Envisions and Challenges in Culinary Co-Creation with GenAI}
\subsubsection{Ongoing Willingness to Exploring ICH Culinary with GenAI}
\label{4.2.1}
After years of working within established culinary frameworks, some chefs reflected that they felt constrained by conventional ways of thinking. GenAI provided them with new perspectives and enabled more creative approaches to cooking. It led participants to express a willingness to innovate and recreate cuisine with the help of GenAI, even looking ahead to future applications, including reinterpretation, customization, and holistic cultural promotion.

After P11 saw the generative results for fusion cuisine, she repeatedly emphasized how the process inspired her. In addition to the fusion with Western cuisine mentioned in the previous session (\ref{4.1.3}), she also planned to explore ideas for co-creating with GenAI to customize Shaanxi Oil-Spilled Noodles, similar to the flexible customization seen in Crossing-the-Bridge Noodles.

GenAI provided P12 with innovative suggestions not only for cooking practices, including new fillings, noodles, and soup bases, but also for promoting the dishes through branding and supporting services. P12 noted, "This is fascinating! It reminded me that innovating and promoting intangible cultural heritage doesn't only revolve around the food itself but should also be explored from multiple dimensions."

\subsubsection{Challenges in Culinary Co-Creation with GenAI}
\label{4.2.2}
For participants who had prior experience with GenAI, they appeared more active and receptive to the workflow. However, their expectations for the quality of generated outputs were also higher. During the workshop, P7 spontaneously requested to use Doubao instead of ChatGPT after a few trials. She said, "I would like to try Doubao for better image generation for Chinese cuisine. I believe it’s better."

P7's feedback reflected the cultural bias present in some GenAI tools. In fact, other participants also encountered similar challenges. For example, P15 noted that the appearance of the Hakka Salt-Baked Chicken designed for youth gatherings (\ref{4.1.1}) deviated too much from traditional Chinese visual styles (Fig. \ref{fig:p15} A). Similarly, GenAI provided P1 with an innovative version of Stuffed Tofu—"Garden-Styled Stuffed Tofu." Although he appreciated the visual appeal, he pointed out that the tomato used was not a variety commonly found in Chinese cuisine (Fig. \ref{fig:p1} B). These biases led participants to perceive the generated images of the co-created recipes as somewhat strange or culturally inconsistent. In this context, Doubao, the Chinese LLM selected by P7, received more positive feedback on image generation than ChatGPT from participants.

Another challenge posed by GenAI tools was the practical feasibility of the generated concepts. P8 pointed out that the Yu-Shiang Shredded Pork Burger proposal mentioned in \ref{4.1.3} faced difficulties during actual execution. Because the dish is a loose stir-fry, it was challenging to use it as a burger filling without losing the integrity of the original sauce. As a result, the concept posed considerable challenges for real-world cooking.

\subsection{Concerns About Losing Traditional Flavors With Culinary Innovation}
\label{4.3}
Despite participants planning to incorporate some of the AI-generated ideas into their future cooking, they also believed not all suggestions from GenAI were valuble to preserve during practice process. The greatest concern among participants was that excessive changes might compromise the essence and flavors of the original dishes. They emphasized that innovation should be done while preserving the core essence of tradition.

For example, P13 expressed: "The reason why ICH dishes have been passed down through generations is because artisans have upheld traditional cooking techniques. Sometimes, the innovative recipes suggested by AI focus too much on novelty and trends, which can easily erode the cultural and traditional features of ICH dishes. While it might attract new tasters, the original diners may no longer experience the flavors that carry local memories and historical heritage."

P7 explored a molecular gastronomy version of Liuyang Steamed Pork Ribs using GenAI (Fig.\ref{fig:p7} C)). However, the AI's suggested sauce (which included lemon juice and honey) strayed too far from the authentic salty, spicy, and savory flavors of the dish. P7 felt that GenAI's understanding of preserving original tastes was insufficient, and she was disappointed that she could not incorporate its suggestion.

\begin{figure*}
    \centering
    \includegraphics[width=1\linewidth]{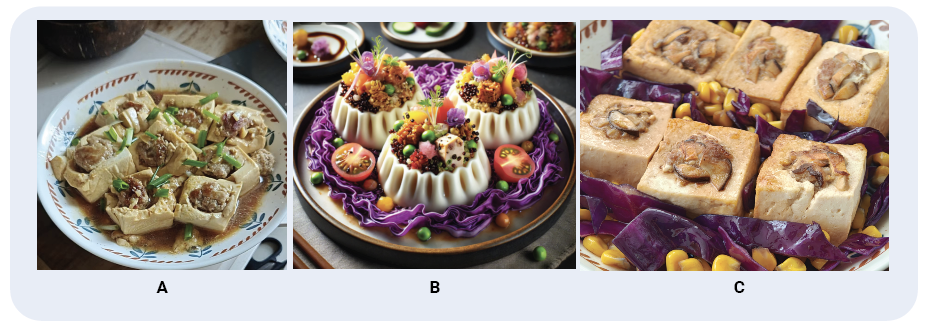}
        \caption{A) Hakka-style Stuffed Tofu prepared by P1; B) Innovated version of Hakka-style Stuffed Tofu generated by GenAI; C) Adjusted version of Hakka-style Stuffed Tofu prepared by P1, based on B) and the related AI-generated recipe.}
    \label{fig:p1}
\end{figure*}

P1 appreciated GenAI's "Garden-Styled Stuffed Tofu" plating suggestion (Fig.\ref{fig:p1} B)), but he did not follow the AI-generated recipe entirely when recreating the dish. He explained: "Following that recipe, the fillings won't be the taste of Hakka Stuffed Tofu. The only adjustment I made was adding some corn to the stuffing." He also noted that he was able to accept the plating suggestion because it did not alter the flavor of the main dish.

Likewise, P15 found many of the recipe innovations interesting, but he was dissatisfied with the GenAI's inability to preserve the original taste of Hakka Salt-Baked Chicken. He commented: "Many of its methods fail to retain the original flavor of Hakka Salt-Baked Chicken. It at best keeps the taste of the salt-baked chicken seasoning powder. But using salt as the baking medium is the soul of this dish. Those suggestions that fail to preserve this essence don't appeal to me."

\begin{figure*}
    \centering
    \includegraphics[width=1\linewidth]{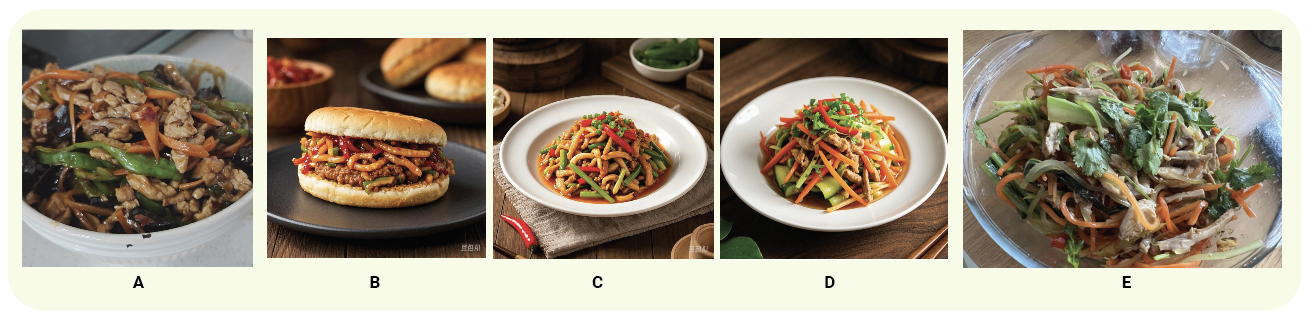}
        \caption{A) Yu-Shiang Shredded Pork prepared by P8; B) Yu-Shiang Shredded Pork Burger innovated by GenAI; C) Vegetarian version of Yu-Shiang Shredded Pork innovated by GenAI; D) Cold appetizer version of Yu-Shiang Shredded Pork innovated by GenAI; E) Cold appetizer version of Yu-Shiang Shredded Pork prepared by P8 based on D) and the related AI-generated recipe.}
    \label{fig:p8}
\end{figure*}

In contrast to the situations above, P8 approved of the innovative recipes generated by GenAI because they preserved the unique sauce flavor of Yu-Shiang Shredded Pork across multiple variations (Fig.\ref{fig:p8}). Whether in the vegetarian recipe or other creative ideas, GenAI ensured that the method retained the proper ratio of two spoons of sugar, two spoons of vinegar, and one spoon of soy sauce. This allowed the innovative dishes to embody the characteristic flavor and essence of Yu-Shiang Shredded Pork. 

\section{Discussion}\label{sec:Discussion}
From the workshop, we found that cuisine practitioners utilized GenAI to reimagine recipes and cooking processes, primarily to promote food traditions to a wider audience. However, they regard the original taste as the essence of culinary culture, which must be preserved in any innovation. In this context, GenAI provided both inspiration and challenges in adapting ICH cuisine for better transmission.

\subsection{Balancing Innovation and Tradition in Cuisine}
Improvements play a important role in safeguarding ICH in globalization \cite{unesco_convention_2003}, especially when adapting local culinary traditions to an international audience \cite{Japan_Washoku}. The cooking practice often requires adjustments to adapt to diverse contexts and situations \cite{cheung2014globalisation}. In our study, we encouraged practitioners to reflect on the development of ICH cuisine and explore recipe recreation using GenAI. Through this process, they attempted to enhance traditional recipes to appeal to a broader group of diners. While practitioners found inspiration in the co-creation process, they also expressed concerns about potential changes in taste that could compromise the essence of their culinary heritage.

To ensure ICH remains representative of its cultural roots, design and innovation must prioritize cultural authenticity and identity \cite{lu_craftsfabrication_2022}. Previous HCI food research has explored ways to address the tendency of technical developments to overlook cultural knowledge in gourmet practices \cite{luo_doufu_2023}. Our findings align with these studies, showing the importance of maintaining a balance between innovation and tradition. While innovation and recreation are valuable for promotion and safeguarding cultural heritage, excessive reinvention, especially when it alters traditional tastes, risks overshadowing preservation and changing the identity of the cuisine.

However, during the innovation process in our study, all participants were experienced chefs, enabling them to identify the key elements of the cuisine that should be preserved, even when GenAI altered some important aspects. For example, P15 emphasized the importance of using "salt" in baking, while P8 appreciated that GenAI retained steps crucial for preserving a key flavor (\ref{4.3}). This suggests that concerns about losing the balance between innovation and tradition may be alleviated if practitioners possess a deep understanding of their cultural heritage.

\subsection{ICH Culinary Creativity with GenAI}
GenAI has the ability to enhance expressive narratives for innovating and reflecting on cultural heritage \cite{fu_being_2024}. According to our findings, using GenAI as a co-creation tool in ICH cuisine innovation can help chefs break traditional thinking and reimagine recipe creation through its capabilities. Particularly when aiming to make the cuisine more widely accepted, the knowledge provided by GenAI offered chefs new ideas, such as incorporating elements from other cuisines or substituting healthier ingredients. The narratives behind these reimagined recipes were not entirely fabricated but were grounded in food traditions, making it easier for chefs to accept and integrate them into their practices.

Research on human collaboration with GenAI in creative tasks \cite{lc_speculative_2024}has shown the potential of AI for fostering creative ideation \cite{han_when_2024}. In this study, the text-to-image capabilities of GenAI further inspired participants to imagine improved cuisines, increasing their interest in co-creating recipes. This was particularly relevant for Chinese cuisine, which emphasizes not only taste and smell but also visual presentation \cite{foods12193628}.

Our findings also revealed that the co-creation process enabled chefs to envision updated recipes with ingredient substitutions and alternative cooking methods, leading to innovative dish variations that transcend traditional forms. This demonstrates that GenAI can serve as a digital tool for inspiring new culinary expressions rooted in traditional food heritage. These findings align with previous research on digital ICH, which highlights the role of digital technologies in preserving ICH by making it both accessible and adaptable to contemporary contexts \cite{liu_digital_2023}. Digital tools should not only document ICH but also support its evolution, enabling sustainability alongside innovation.

\subsection{Limitations of Using GenAI in ICH Recreation}
While GenAI proved useful in aiding cuisine innovation and generating new ideas for recipe adaptation, several challenges emerged during its application in ICH recreation. One significant challenge was the limited contextual knowledge of GenAI regarding ICH cuisine. The knowledge and techniques of ICH culinary traditions are often transmitted through oral traditions or handwritten notes, which are easily lost over time \cite{unesco_practices_2023, frey_preserving_2015, dietz_preserving_2017}. Due to the lack of transmission materials, GenAI models sometimes struggled to provide accurate and relevant suggestions for cooking practices. For example, although GenAI provided an interesting idea to P8 by integrating Yu-Shiang Shredded Pork into a burger (\ref{4.2.2}), it failed to realize that the dish’s texture is somewhat fragmented, making it difficult to function as a burger ingredient. This limitation occasionally hindered the innovation process, as the resulting recipes were not always practical.

Another challenge was the cultural bias embedded in the GenAI model. The generated outputs occasionally suggested kitchen utensils or ingredients that were heavily influenced by Western culinary traditions \cite{Typology_of_Risks}. In \ref{4.2.2}, the examples of P1 and P15 illustrated that the ChatGPT generated  food images also reflected bias shaped by Western culinary culture. Food heritage is especially a part of cultural identity, and this caused the innovation with cultural bias to be unacceptable to participants. The mismatch between the Western model’s outputs and non-Western culinary contexts highlights the limitations of the model, and underscores the importance of integrating more inclusive and diverse training data. Another approach would be to apply a locally trained model for projects with a strong cultural identity.

Lastly, GenAI faced limitations in addressing the sensory aspects of food innovation. Taste and smell are critical elements of cuisine, but GenAI lacks the capability to engage with these sensory experiences. Its suggestions were based entirely on theoretical knowledge rather than practical or sensory-driven insights, making it difficult for participants to fully trust or incorporate the tool into sensory-dependent aspects of ICH cuisine innovation. This highlights the need for further development in training LLMs with more sensory data before applying them to food heritage-related projects.

\subsection{Limitations}
\subsubsection{Selection of Participants}
The study primarily included experienced chefs and long-term practitioners of Chinese cuisine residing abroad. While their expertise contributed valuable insights, the diverse and highly regional nature of Chinese cuisine was not fully captured. Chinese cuisine varies significantly across regions, with distinct ingredients, techniques, and cultural significance. By focusing on participants who cook a variety of Chinese cuisines case by case, the study may not have fully represented chefs with specialized knowledge of specific regional cuisines. To address this, future studies could recruit participants with expertise in a particular regional cuisine to ensure a deeper understanding of how innovation and preservation are approached within that specific context.

\subsubsection{Lack of Evaluation on Tastes}
Taste is a critical component of cuisine, especially in the context of ICH. In this study, only the researchers and chefs in the workshop tasted the original and some of the innovative dishes. This limited evaluation of taste makes it difficult to objectively assess whether the innovations were successful from a culinary perspective. To improve upon this, future research could involve professional tasters or a diverse panel of diners to evaluate and compare the original and innovative cuisines. This would provide a broader and more objective assessment of the culinary outcomes.

\subsubsection{Lack of Evaluation on ICH Characteristics}
The evaluation of whether the innovative recipes retained the characteristics of ICH cuisine was conducted only by the chefs themselves. While their perspectives are valuable, the absence of external evaluations limits the depth of the findings regarding cultural authenticity. Future research could involve professionals in the fields of cultural heritage, culinary arts, or gastronomy to provide a more comprehensive and critical evaluation of authenticity. Additionally, facilitated group discussions or panels involving both chefs and cultural experts could further enrich the understanding of whether the innovated recipes align with ICH principles.

\section{Conclusion}\label{sec:Conclusion}
This study demonstrates the potential of GenAI as a tool for supporting innovation and adaptation in Chinese cuisine as ICH. Through workshops involving 15 experienced practitioners of traditional Chinese cuisine, participating chefs co-created recipes with GenAI, using traditional cuisine as a foundation to develop innovative dishes that appeal to broader audiences. The findings showcase how GenAI can help reimagine heritage cuisines to meet contemporary tastes while maintaining their cultural relavance. However, the study also reveals the importance of balancing innovation and preservation. While adapting and modernizing recipes is crucial for safeguarding ICH cuisine in a globalized world, excessive innovation may harm traditional flavors and cultural essence. This tension between innovation and cultural authenticity calls for careful consideration in how GenAI and similar technologies are utilized. Overall, this study provides important implications, emphasizing the need to integrate such tools thoughtfully to ensure ICH food evolves in meaningful ways without compromising its traditions.

\begin{acks}
This research was supported by the TRS Theme-based Research Scheme (T45-205/21-N) (CityU Project No: 8779030): Building Platform Technologies for Symbiotic Creativity in Hong Kong.

\end{acks}

\bibliographystyle{ACM-Reference-Format}
\bibliography{references}

\end{document}